# Extracting single crystal diffraction pattern from powder diffraction by intensity correlation functions

Yun Zhao

Department of Physics, Arizona State University, Tempe, Arizona 85287, USA

**Abstract**

We applied the analysis of x-ray intensity angular correlation function to dilute ensembles of identical spinel crystals. Firstly, we show that the angular correlation from measured diffraction patterns with many crystals per shot converges to the correlation for single crystal. Secondly, we determined the diffraction pattern from single crystal from diffraction patterns of many crystals. Finally, we discussed the full 3D reciprocal space retrieval and its potential application in solving structure with dilute powder diffraction data, where the intensity fluctuation could be observed on each ring.

## 1  Introduction

Rietveld refinement is a powerful approach to determine structure of crystals from powder diffraction data. Many programs available online have been developed based on this approach.[1] However, the success of this approach requires a good model first. In order to collect powder diffraction data, sample of small crystals have to be exposed to X-ray for long period of time, which may introduce significant radiation damage. Kam first pointed out that the three-dimensional structure of one particle may be determined using the x-ray scattering from many randomly oriented copies, without modeling of priori information[2][3]. Meanwhile, it was shown that the signal to noise ratio is the same for single particle and multiply particles per shot[4]. However, this method is undeveloped for about 20 years after Kam' paper as lack of brief and intense x-ray sources. With the availability of free electron laser, this idea was re-discovered and much theoretical work are under developing. Saldin et al did much proof on principle simulations in single particle structure determination as well as experiments[5][6][7].

In this report, we focus on the application of this method on crystal structure determination. Because the ensemble of crystals are static throughout the snapshot exposure, spinel crystals scattering patterns contain angular intensity fluctuations and thus differ from conventional powder diffraction pattern. These intensity fluctuations may provide us additional information on structure determination. It will be shown that the diffraction pattern for single crystal can be recovered by fluctuation pair and triple correlation functions alone, without other priori information.

## 2  Theory
2.1  Spinel powder diffraction

For the coherent monochromatic plane wave, the incident and outgoing wavevector

can be denoted as $\vec{k}_i$ and $\vec{k}_o$. The structure factor for a unit cell is given by

$$F\_cell(\vec{q}) = \sum_i f_i \exp(i * \vec{q} * \vec{r_i})$$

where $\vec{q} = \vec{k}_o - \vec{k}_i$, $\vec{r_i}$ is the atomic coordinates in unit cell, $f_i$ is the corresponding atomic scattering factor.

The structure factor for lattice is given by

$$F\_lattice = \sum_n \exp(i * \vec{q} * \vec{r_n})$$

where $\vec{r_n}$ is the displacement of the nth unit cell with respect to origin one. It will converge to a delta function as crystal goes to infinite.

Then, the scattering intensity from one crystal is

$$I(\vec{q}) \propto |F_{xtal}|^2 = \left|\sum_i f_i \exp(-\vec{q}*\vec{r_i}) \sum_n \exp(-\vec{q}*\vec{r_n})\right|^2$$

Here we assume that different crystals scatter x-ray incoherently. Thus, the intensity observed on detector is simply the sum of the intensity from individual crystal.

$$I_k(\vec{q}) = \sum_i^{N_c} I(\vec{q}, \omega_k^i)$$

where $\omega_k^i$ is the orientation of i-th crystal during k-th snapshot. $N_c$ is the number of crystals illuminated during k-th diffraction pattern. Because the number of crystals in correlated x-ray scattering is much less than that in conventional powder diffraction, we may observe the spotty rings which reflects intensity fluctuations.

2.2 angular correlation functions

For the diffraction pattern of a single crystal, the pair correlation function for two different rings is defined as

$$C_1(q_i, q_j, \Delta\varphi) = \frac{1}{N_\varphi} \sum_m^{N_\varphi} I(q_i, \varphi_m) I(q_j, \varphi_m + \Delta\varphi) \qquad (1)$$

where $q_i$ and $q_j$ represents radius of the i-th and j-th ring on diffraction pattern..

$N_\varphi$ is the number of azimuthal angels at $\varphi_m$ which the intensity are measured. In a similar way, the triple correlation function is defined as

$$T_1(q_i, q_j, \Delta\varphi) = \frac{1}{N_\varphi} \sum_m^{N_\varphi} I(q_i, \varphi_m)^2 I(q_j, \varphi_m + \Delta\varphi) \qquad (2)$$

For many crystals case, the fluctuation pair correlation, which could be directly calculated from experimental data, is defined as

$$C_{exp} = \langle \left(\sum_l^{N_c} I(q_i, \omega_k^l) - <I_k(q_i)>_k\right)\left(\sum_m^{N_c} I(q_j, \omega_k^m) - <I_k(q_j)>_k\right)\rangle_k \qquad (3)$$

Then the pair correlation function for single crystal can be extracted by

$$C_1(q_i, q_j, \Delta\varphi) = \frac{1}{N_c} C_{exp} + \frac{1}{N_c^2} <I_k(q_i)>_k^2 \qquad (4)$$

In a similar fashion, the fluctuation triple correlation function is defined as

$$T_{exp} = \langle \left(\sum_l^{N_c} I(q_i, \omega_k^l) - <I_k(q_i)>_k\right)^2 \left(\sum_m^{N_c} I(q_j, \omega_k^m) - <I_k(q_j)>_k\right)\rangle_k \qquad (5)$$

Then the triple correlation function for single crystal can be extracted by

$$T_1(q_i, q_j, \Delta\varphi) = \frac{1}{N} T_{exp}(q_i, q_j) + \frac{2}{N} \langle I_k(q_j)\rangle_k C_1(q_i, q_j) - \frac{1}{N^3} \langle I_k(q_i)\rangle_k^2 \langle I_k(q_j)\rangle_k \qquad (6)$$

2.3 reconstruction of single particle diffraction pattern

The intensity of diffraction pattern can be expanded in circular harmonics

$$I(q, \varphi) = \sum_m I_m(q) \exp(im\varphi) \qquad (7)$$

In general, $I_m(q)$ are complex numbers. Taking the Fourier transform of $C_1(q_i, q_j, \Delta\varphi)$ and $T_1(q_i, q_j, \Delta\varphi)$, we have

$$B_m(q_i, q_j) = \frac{1}{N_\varphi} \sum_{\Delta\varphi} C_1(q_i, q_j, \Delta\varphi) \exp(-im\Delta\varphi)$$

$$= I_m(q_i) I_m^*(q_j) \qquad (8)$$

$$FT_m^{(obs)}(q_i, q_j) = \frac{1}{N_\varphi} \sum_{\Delta\varphi} T_1(q_i, q_j, \Delta\varphi) \exp(-im\Delta\varphi) \qquad (9)$$

and it can be shown that

$$FT_m^{(calc)}(q_i, q_i) = I_m^*(q_i) \sum_{M \neq 0, m} I_M(q_i) I_{m-M}(q_i) \qquad \text{for } m \neq 0. \qquad (10)$$

So the magnitude of $I_m(q_i)$ is determined by $|I_m(q_i)| = \sqrt{B_m(q_i, q_i)}$. The unknown phases needs to be determined to reconstruct the single crystal diffraction pattern.

## 3 Application to spinel powder diffraction

Each spinel crystal has 10 unit cells in x and y direction, with lattice constant

8.0858 Å.. The wavelength of incoming x-rays is 1.5406 Å. Flat Ewald sphere is assumed in the present simulation. The simulated diffraction pattern from single crystal is shown as follow

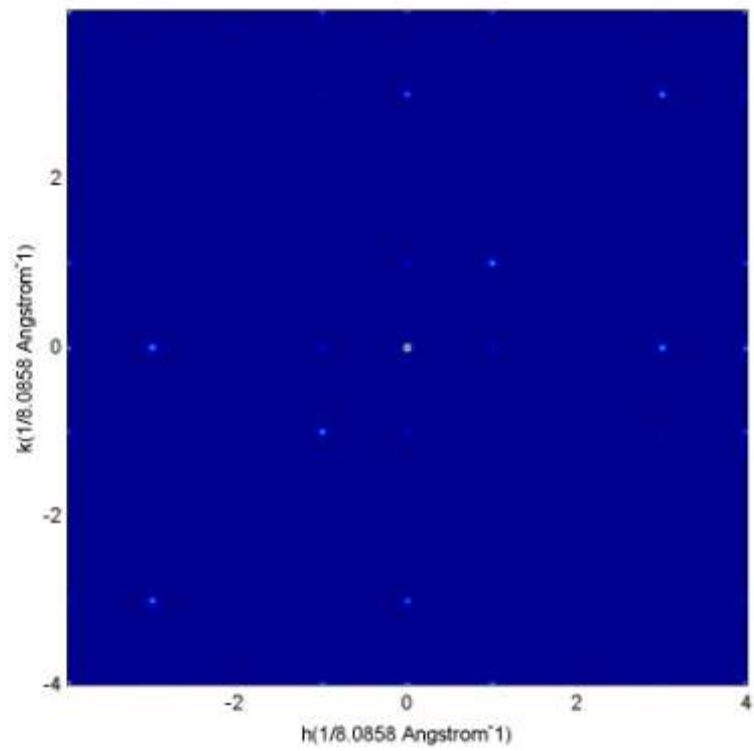

Fig 1    Diffraction pattern for single crystal

Next we simulate powder diffractions where 10 crystals are illuminated simultaneously per shot. Each crystal lies in a random orientation along z axis and scatters x-ray incoherently. In this way, we may obtain spotty powder diffraction rings, as shown in Fig 2.

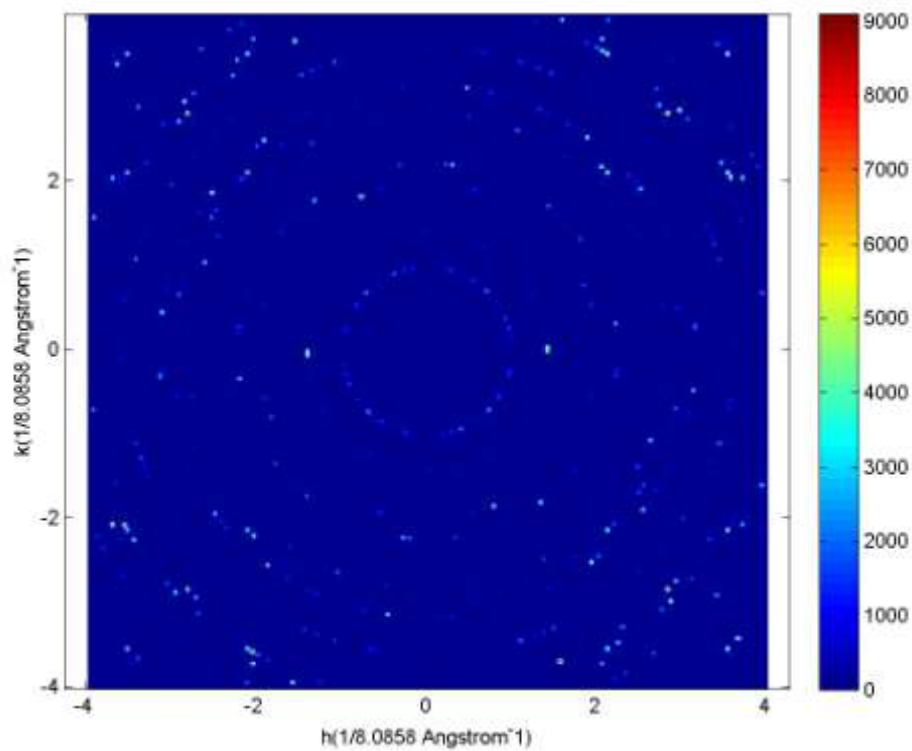

Fig 2 Diffraction pattern for 10 crystals

In this report, we mainly investigate whether we could recover the diffraction pattern for single crystal (Fig 1) from powder diffraction data (Fig 2) . First, we need to obtain convergent values for angular correlation functions by averaging them over a large number of multiple-crystal diffraction patterns. In my case, 100 diffraction patterns were simulated. The averaged angular autocorrelation function shows the convergence to single crystal (Fig 3).

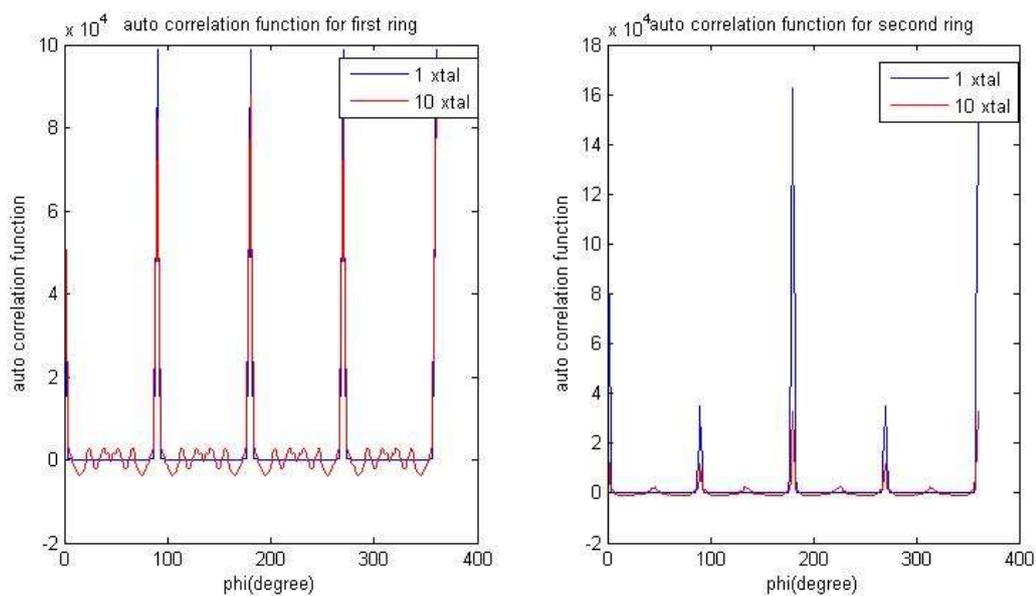

Fig 3  Convergence of auto correlation functions for first and second ring

The magnitude of $I_m(q_i)$ can be uniquely determined by taking the square root of $B_m(q_i, q_i)$. Its phase could be solved by the charge-flipping method described in [8]. In present report, we take all $I_m(q_i)$ to be real and maximum value of m is 38. Note that $I_{-m}(q_i) = I_m^*(q_i)$ as a results of Friedel's rule. So only even values are non-zero. Here we take all coefficients as real. Only signs need to be determined. After searching $2^{19}$ combinations of signs to optimize the function.

$$\sum_{m \neq 0} |FT_m^{(obs)} - FT_m^{(calc)}|^2$$

The result of reconstructing the single diffraction pattern is shown in Fig 4.

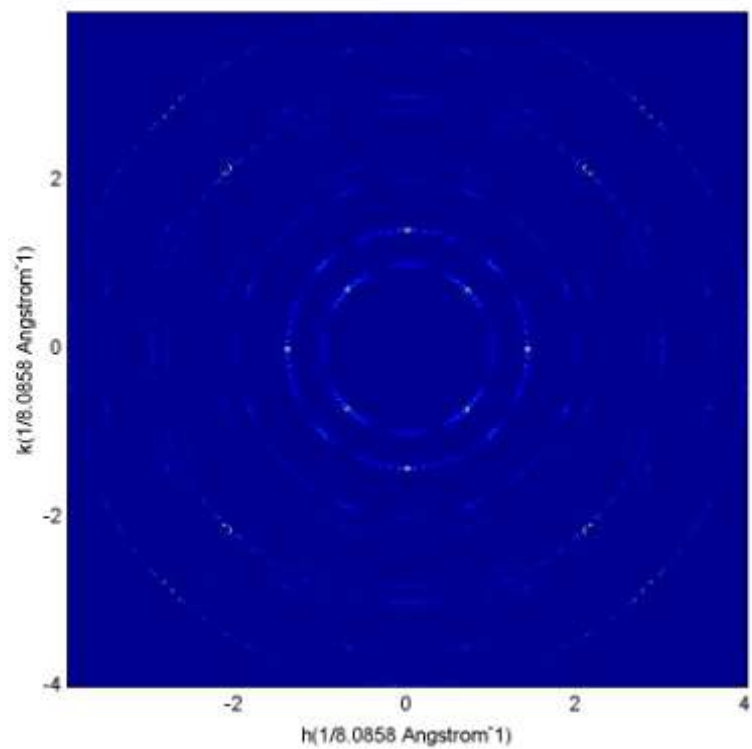

Fig 4  Single crystal diffraction pattern reconstructed from the magnitude of $I_m(q_i)$ determined from the mean pair correlation $C_1$, and signs from the mean triple correlations $T_1$ from 100 multi-particle diffraction pattern like that of figure 2.

**4  3D structure determination**

So far, we reconstructed the 2D low resolution diffraction pattern. Much effort are still required to develop this method to 3D reconstruction before real application in powder diffraction experiment. Firstly, we cannot obtain powder diffraction pattern just by rotating crystal along one axis in real experiment. All orientations need to be adequately sampled. Secondly, we should note that the diffraction pattern reconstructed is low resolution data. In order to get high resolution data, the diffraction pattern will be the projection from curved Ewald sphere.

According to my knowledge, no simulation or experiment work on real 3D structure reconstruction by this method has been done. The low resolution diffraction pattern probably origins from the assumption of flat Ewald sphere. Elser generated this method to a semi 3D case [7], where particles can be aligned in random orientation on a 2D substrate which can tilt freely with respect to the x-ray beam. As the tilt angle between substrates and x-ray beam can be measured and the correlation function has the same property as eqn (8) and (10), the reconstruction proceeds pretty much the same as for 2D case [9].

For the full rotation freedom case, the reconstruction idea is still the same. Firstly, we need to obtain convergent pair and triple correlation functions from simulated powder diffraction patterns. Then we expand the 3D reciprocal-space map by spherical harmonics [10].

$$I(\vec{q}) = \sum_{lm} I_{lm}(\vec{q}) Y_{lm}(\vec{q})$$

It can be shown that

$$C_1(q_i, q_j, \Delta\varphi) = \frac{1}{4\pi} \sum_{l}^{l_{max}} P_l(\cos[\Delta\phi]) B_l(q_i, q_j)$$

where $C_1(q_i, q_j, \Delta\varphi)$ is the ring cross correlation, $P_l(\cos[\Delta\phi])$ are the Legendre polynomials, and

$$B_l(q_i, q_j) = \sum_{m=-l}^{l} I_{lm}(q_i) I_{lm}^*(q_j)$$

Then we need to find all the complex coefficients involved from the above equation. It is a formidable task either using triple correlation method or phase iterative method.[11]

## 6  Discussion and Conclusion

Here we mainly demonstrate that the 2D diffraction pattern from single crystal can be reconstructed from powder diffraction data, in principle. There are still several limits on the present 2D simulation. First, we may observe that the intensity of (110) spot is not equivalent to ($\bar{1}10$) from the single particle diffraction pattern. But the reconstructed diffraction pattern couldn't distinguish this pair. From the pair correlation function, we could observe the intensity variation. it seems that this inefficiency doesn't origin from the expansion order $m_{max}$, but the accuracy of phase where assume all coefficient to be real. Secondly, a proper reference ring is crucial for successful reconstruction both in triple correlation method or phase iterative method. In this report, we chose the first ring as our reference ring and then calculated the pair correlation function with respect to the first ring, which indicates the relative position information of Bragg spots on different rings. The phases of high resolution rings are not well recovered. It may be improved by choosing several outer rings as reference ring[8].

Although the diffraction pattern reconstruction demonstration in this report is two dimensional, this idea provides us much insights on the application of real 3D powder diffraction. As to the 3D diffraction volume reconstruction, much effort on theory are still needed to develop.

**Acknowledgement**
The author thanks Prof. John Spence for discussion.

# Appendix I: Intensity angular correlation functions

## 0 Notation:

$I(\mathbf{q_i}, \omega_k^l)$        the intensity contribution from the l_th crystal during k_th snapshot, with its orientation $\omega_k^l$, scattering vector $\mathbf{q_i}$ (Bold letter means a vector, letters without bold mean magnitude).

$I_k(\mathbf{q_i})$        the observed intensity from k_th snapshot

**Note:** In the following of this report, I without subscript k always means the diffraction intensity from one crystal. $I_k$ always represents the observed diffraction intensity which results from x-ray scattered by many crystals.

$\tilde{I}_k(\mathbf{q_i})$        fluctuation intensity from k_th snapshot

$C_1(\mathbf{q_i}, \mathbf{q_j}, \Delta\varphi)$        angular pair correlation function for single crystal

$T_1(\mathbf{q_i}, \mathbf{q_j}, \Delta\varphi)$        angular triple correlation function for single crystal

$\tilde{C}_{exp}(\mathbf{q_i}, \mathbf{q_j}, \Delta\varphi)$        fluctuation angular pair correlation function for multiple crystals, which is averaged over all experimental or simulated powder diffraction patterns.

$\tilde{T}_{exp}(\mathbf{q_i}, \mathbf{q_j}, \Delta\varphi)$        fluctuation angular triple correlation function for multiple crystals, which is averaged over all experimental or simulated powder diffraction patterns.

$B_m(\mathbf{q_i}, \mathbf{q_j})$        Fourier transform of $C_1(\mathbf{q_i}, \mathbf{q_j}, \Delta\varphi)$

$FT(\mathbf{q_i}, \mathbf{q_j}, \Delta\varphi)$        Fourier transform of $T_1(\mathbf{q_i}, \mathbf{q_j}, \Delta\varphi)$

## 1 Obtain $C_1(\mathbf{q_i}, \mathbf{q_j}, \Delta\varphi)$ from $C_{exp}(\mathbf{q_i}, \mathbf{q_j}, \Delta\varphi)$

For the diffraction pattern from a single crystal, the pair correlation function for two different rings is defined as

$$C_1(q_i, q_j, \Delta\varphi) = \frac{1}{N_\varphi} \sum_{m}^{N_\varphi} I(q_i, \varphi_m) I(q_j, \varphi_m + \Delta\varphi) \qquad (1)$$

where $q_i$ and $q_j$ represents radius of the i-th and j-th ring on diffraction pattern.. $N_\varphi$ is the number of azimuthal angels at $\varphi_m$ which the intensity are measured. In a similar way, the triple correlation function is defined as

$$T_1(q_i, q_j, \Delta\varphi) = \frac{1}{N_\varphi} \sum_{m}^{N_\varphi} I(q_i, \varphi_m)^2 I(q_j, \varphi_m + \Delta\varphi) \qquad (2)$$

Now let's consider many-crystal case. Here we assume each crystal scatters x-ray incoherently, thus the intensity observed on detector is simply the sum of the intensity from each individual crystal.

$$I_k(\mathbf{q_i}) = \sum_l^{N_c} I(\mathbf{q_i}, \omega_k^l)$$

where $\omega_k^l$ is the orientation of l_th crystal during k_th snapshot.

The fluctuation intensity is defined as

$$\tilde{I}_k(\mathbf{q_i}) = I_k(\mathbf{q_i}) - <I_k(\mathbf{q_i})>_k$$

where the second term means average over all diffraction patterns

$$<I_k(\mathbf{q_i})>_k = \frac{1}{N_d} \sum_{k=1}^{N_d} I_k(\mathbf{q_i})$$

where $N_d$ is the total number of diffraction patterns.
The fluctuation pair correlation for simulated diffraction patterns is defined as

$$\tilde{C}_{exp}(q_i, q_j, \Delta\varphi) = \frac{1}{N_d} \sum_k \frac{1}{2\pi} \int_0^{2\pi} \tilde{I}_k(q_i, \varphi) \tilde{I}_k(q_j, \varphi + \Delta\varphi) \, d\varphi$$

Let's change the integral by sum,

$$\tilde{C}_{exp}(q_i, q_j, \Delta\varphi) = \frac{1}{N_\varphi} \frac{1}{N_d} \sum_m^{N_\varphi} \sum_k^{N_d} \tilde{I}_k(q_i, \varphi_m) \tilde{I}_k(q_j, \varphi_m + \Delta\varphi)$$

$$= \frac{1}{N_\varphi} \frac{1}{N_d} \sum_m^{N_\varphi} \sum_k^{N_d} \left( \sum_l^{N_c} I(q_i, \varphi_m, \omega_k^l) - <I_k(q_i)>_k \right) \left( \sum_n^{N_c} I(q_j, \varphi_m + \Delta\varphi, \omega_k^n) \right.$$

$$\left. - <I_k(q_j)>_k \right)$$

$$= \frac{1}{N_\varphi} \frac{1}{N_d} \sum_m^{N_\varphi} \sum_k^{N_d} \left( \sum_l^{N_c} I(q_i, \varphi_m, \omega_k^l) \sum_n^{N_c} I(q_j, \varphi_m + \Delta\varphi, \omega_k^n) \right.$$

$$- <I_k(q_i)>_k \sum_n^{N_c} I(q_j, \varphi_m + \Delta\varphi, \omega_k^m)$$

$$\left. - <I_k(q_j)>_k \sum_l^{N_c} I(q_i, \varphi_m, \omega_k^l) + <I_k(q_i)>_k <I_k(q_j)>_k \right)$$

$$= \frac{1}{N_\varphi} \frac{1}{N_d} \sum_m^{N_\varphi} \sum_k^{N_d} \left( \sum_{l=n}^{N_c} I(q_i, \varphi_m, \omega_k^l) I(q_j, \varphi_m + \Delta\varphi, \omega_k^l) \right.$$

$$+ \sum_{l \neq n}^{N_c} I(q_i, \varphi_m, \omega_k^l) I(q_j, \varphi_m + \Delta\varphi, \omega_k^n)$$

$$- <I_k(q_i)>_k \sum_l^{N_c} I(q_j, \varphi_m + \Delta\varphi, \omega_k^l)$$

$$\left. - <I_k(q_j)>_k \sum_l^{N_c} I(q_i, \varphi_m, \omega_k^l) \right) + <I_k(q_i)>_k <I_k(q_j)>_k$$

Note that in the first term, $\frac{1}{N_\varphi} \sum_m^{N_\varphi} I(q_i, \varphi_m, \omega_k^l) I(q_j, \varphi_m + \Delta\varphi, \omega_k^l)$, is the pair correlation function $C_1(q_i, q_j, \Delta\varphi)$ that would arise from single crystal. The second uncorrelated term can be expressed as

$$\sum_{l \neq n}^{N_c} \left( \frac{1}{N_\varphi} \frac{1}{N_d} \sum_m^{N_\varphi} \sum_k^{N_d} I(q_i, \varphi_m, \omega_k^l) \right) * \left( \frac{1}{N_\varphi} \frac{1}{N_d} \sum_m^{N_\varphi} \sum_k^{N_d} I(q_j, \varphi_m + \Delta\varphi, \omega_k^n) \right)$$

When we take average over all diffraction pattern and integrate over each ring, the relative orientation between crystals $\omega$ will be washed out. And both terms above will not depend on angle or any specific crystal. Here I simply denoted them as $\langle I(q_i) \rangle_\omega$ and $\langle I(q_j) \rangle_\omega$. As the sum $\sum_{l \neq n}^{N_c}$ has $(N_c^2 - N_c)$ terms. Hence, the second term has the following simple expression

$$= (N_c^2 - N_c) \langle I(q_i) \rangle_\omega \langle I(q_j) \rangle_\omega$$

And

$$\tilde{C}_{exp}(q_i, q_j, \Delta\varphi)$$

$$= N_c C_1(q_i, q_j, \Delta\varphi) + (N_c^2 - N_c) \langle I(q_i) \rangle_\omega \langle I(q_j) \rangle_\omega$$

$$- N_c <I_k(q_i)>_k \langle I(q_j) \rangle_\omega - N_c \langle I(q_i) \rangle_\omega <I_k(q_j)>_k$$

$$+ <I_k(q_i)>_k <I_k(q_j)>_k$$

Note that $\langle I(q) \rangle_\omega$ is the average intensity from single crystal, $<I_k(q)>_k$ is the average intensity from $N_c$ crystals during a x-ray shot. And $\langle I(q) \rangle_\omega$, $<I_k(q)>_k$ are uniform on each ring. So $\langle I(q) \rangle_\omega = \frac{1}{N_c} <I_k(q)>_k$. Then

$$\tilde{C}_{exp}(q_i, q_j, \Delta\varphi)$$
$$= N_c C_1(q_i, q_j, \Delta\varphi) + (1 - \frac{1}{N_c} - 1 - 1 + 1) <I_k(q_i)>_k <I_k(q_j)>_k$$

Hence,
$$C_1(q_i, q_j, \Delta\varphi) = \frac{1}{N_c}\tilde{C}_{exp}(q_i, q_j, \Delta\varphi) + \frac{1}{N_c^2} <I_k(q_i)>_k <I_k(q_j)>_k$$

In the above equation, $\tilde{C}_{exp}(q_i, q_j, \Delta\varphi)$ and $<I_k(q_i)>_k$ can be easily calculated from diffraction patterns by definition. Hence, pair correlation for single crystal $C_1(q_i, q_j, \Delta\varphi)$ is solved.

## 2 derivation for triple angular correlation

The fluctuation triple correlation for simulated diffraction patterns is defined as

$$\tilde{T}_{exp}(q_i, q_j, \Delta\varphi) = \frac{1}{N_\varphi}\frac{1}{N_d}\sum_{m}^{N_\varphi}\sum_{k}^{N_d} \tilde{I}_k^2(q_i, \varphi_m)\tilde{I}_k(q_j, \varphi_m + \Delta\varphi)$$

This term can be directly calculated from all the diffraction patterns. In a similar fashion, $\tilde{T}_{exp}(q_i, q_j, \Delta\varphi)$ can be expanded as

$$= \frac{1}{N_\varphi}\frac{1}{N_d}\sum_{m}^{N_\varphi}\sum_{k}^{N_d}\left(\left(\sum_{l}^{N_c} I_k(q_i, \omega_k^l)\right)^2 - 2*\sum_{l}^{N_c} I_k(q_i, \omega_k^l) <I_k(q_i)>_k\right.$$
$$\left. +<I_k(q_i)>_k^2\right)\left(\sum_{n}^{N_c} I_k(q_j, \omega_k^n) - <I_k(q_j)>_k\right)$$

$$= N_c T_1(q_i, q_j, \Delta\varphi) + (N_c^2 - N_c)\langle I_k(q_i)^2\rangle_\omega \langle I_k(q_j)\rangle_\omega - 2N_c \langle I_k(q_i)\rangle_k C_1(q_i, q_j, \Delta\varphi)$$

$$-(N_c^2 - N_c)\langle I_k(q_i)\rangle_k \langle I_k(q_i)\rangle_\omega \langle I_k(q_j)\rangle_\omega - N_c^2 \langle I_k(q_i)\rangle_k^2 \langle I_k(q_j, \omega_k^n)\rangle_\omega$$

$$- N_c^2 \langle I_k(q_i)^2\rangle_\omega \langle I_k(q_j, \omega_k^n)\rangle_k + 2N_c^2 \langle I_k(q_i)\rangle_k \langle I_k(q_j)\rangle_k \langle I_k(q_i)\rangle_\omega$$

$$+ N_c^2 \langle I_k(q_i)\rangle_k^2 \langle I_k(q_j)\rangle_k$$

$$= N_c T_1(q_i, q_j, \Delta\varphi) - 2N_c \langle I_k(q_i)\rangle_k C_1(q_i, q_j, \Delta\varphi) + [N_c^2 - N_c - 2(N_c^2 - N_c) - N_c^2$$
$$- N_c^2 + 2N_c^2 + N_c^2]\, \langle I_k(q_i)\rangle_k^2 \langle I_k(q_j)\rangle_k$$

$$= N_c T_1(q_i, q_j, \Delta\varphi) - 2N_c \langle I_k(q_i)\rangle_k C_1(q_i, q_j, \Delta\varphi) + N_c \langle I_k(q_i)\rangle_k^2 \langle I_k(q_j)\rangle_k$$

So

$$T_1(q_i, q_j, \Delta\varphi) = \frac{1}{N_c}\tilde{T}_{exp}(q_i, q_j, \Delta\varphi) + \frac{2}{N_c}\langle I_k(q_j)\rangle_k C_1(q_i, q_j, \Delta\varphi)$$
$$- \frac{1}{N_c^3}\langle I_k(q_i)\rangle_k^2\langle I_k(q_j)\rangle_k$$

All terms on the right side of equation can be calculated from diffraction patterns, hence triple correlation for single crystal is solved.

## 3  Fourier transform of pair angular correlation

By definition, the Fourier transform of pair angular correlation is given by

$$B_m(q_i, q_j) = \frac{1}{2\pi}\int_0^{2\pi} C_1(q_i, q_j, \Delta\varphi)\exp(-im\Delta\varphi)\, d\Delta\varphi \qquad (1)$$

where

$$C_1(q_i, q_j, \Delta\varphi) = \frac{1}{2\pi}\int_0^{2\pi} I(q_i, \varphi_m)I(q_j, \varphi_m + \Delta\varphi)\, d\varphi$$

Expand $I(q, \varphi)$ in circular harmonics, then

$$C_1(q_i, q_j, \Delta\varphi) = \frac{1}{2\pi}\int_0^{2\pi}\left(\sum_n I_n(q_i)\exp(in\varphi)\right)\left(\sum_l I_l(q_j)\exp[il(\varphi + \Delta\varphi)]\right) d\varphi$$

$$= \frac{1}{2\pi}\int_0^{2\pi}\sum_n\sum_l I_n(q_i)I_l(q_j)\exp(in\varphi)\exp[il(\varphi + \Delta\varphi)]\, d\varphi \qquad (2)$$

Put eqn (2) to (1)

$$B_m(q_i, q_j) = \left(\frac{1}{2\pi}\right)^2 \int_0^{2\pi}\int_0^{2\pi}\sum_n\sum_l I_n(q_i)I_l(q_j)\exp(in\varphi)\exp[il(\varphi + \Delta\varphi)]\exp(-im\Delta\varphi)\, d\varphi\, d\Delta\varphi$$

$$= \left(\frac{1}{2\pi}\right)^2 \int_0^{2\pi}\int_0^{2\pi}\sum_n\sum_l I_n(q_i)I_l(q_j)\exp[i(n+l)\varphi]\exp[i(l-m)\Delta\varphi]\, d\varphi\, d\Delta\varphi$$

Note that

$$\frac{1}{2\pi}\int_0^{2\pi}\exp[i(l-m)\Delta\varphi]\, d\Delta\varphi = \begin{cases} 1 & \text{when } l = m \\ 0 & \text{when } l \neq m \end{cases}$$

Hence

$$B_m(q_i, q_j) = \frac{1}{2\pi}\int_0^{2\pi}\sum_n I_n(q_i)I_m(q_j)\exp[i(n+m)\varphi]\, d\varphi$$

Also note that

$$\frac{1}{2\pi}\int_0^{2\pi}\exp[i(n+m)\varphi]\, d\varphi = \begin{cases} 1 & \text{when } n = -m \\ 0 & \text{when } n \neq -m \end{cases}$$

So

$$B_m(q_i, q_j) = I_{-m}(q_i)I_m(q_j)$$

As $I_{-m}(q_i) = I_m^*(q_i)$, so

$$B_m(q_i, q_j) = I_m^*(q_i)I_m(q_j)$$

## 4 Fourier transform of triple angular correlation

By definition, the Fourier transform of triple angular correlation is given by

$$FT_m^{(obs)}(q_i, q_j) = \frac{1}{2\pi}\int_0^{2\pi} T_1(q_i, q_j, \Delta\varphi) \exp(-im\Delta\varphi)\, d\Delta\varphi$$

or

$$FT_m^{(obs)}(q_i, q_j) = \frac{1}{N_\varphi}\sum_{\Delta\varphi} T_1(q_i, q_j, \Delta\varphi) \exp(-im\Delta\varphi)$$

where

$$T_1(q_i, q_j, \Delta\varphi) = \frac{1}{2\pi}\int_0^{2\pi} I(q_i, \varphi_m)^2 I(q_j, \varphi_m + \Delta\varphi)\, d\varphi$$

Expand $I(q, \varphi)$ in circular harmonics, then

$$T_1(q_i, q_j, \Delta\varphi) = \frac{1}{2\pi}\int_0^{2\pi} \left(\sum_n I_n(q_i)\exp(in\varphi)\right)^2 \left(\sum_l I_l(q_j)\exp[il(\varphi + \Delta\varphi)]\right) d\varphi$$

$$= \frac{1}{2\pi}\int_0^{2\pi} \sum_{n_1}\sum_{n_2}\sum_l I_{n_1}(q_i)I_{n_2}(q_i)I_l(q_j)\exp(in_1\varphi)\exp(in_2\varphi)\exp[il(\varphi + \Delta\varphi)]\, d\varphi \quad (2)$$

Put eqn (2) to (1)

$$FT_m(q_i, q_j)$$

$$= \left(\frac{1}{2\pi}\right)^2 \int_0^{2\pi}\int_0^{2\pi} \sum_{n_1}\sum_{n_2}\sum_l I_{n_1}(q_i)I_{n_2}(q_i)I_l(q_j)\exp(in_1\varphi)\exp(in_2\varphi)\exp[il(\varphi + \Delta\varphi)]\exp(-im\Delta\varphi)\, d\varphi\, d\Delta\varphi$$

$$= \left(\frac{1}{2\pi}\right)^2 \int_0^{2\pi}\int_0^{2\pi} \sum_{n_1}\sum_{n_2}\sum_l I_{n_1}(q_i)I_{n_2}(q_i)I_l(q_j)\exp[i(n_1 + n_2 + l)\varphi]\exp[i(l - m)\Delta\varphi]\, d\varphi\, d\Delta\varphi$$

Note that

$$\frac{1}{2\pi}\int_0^{2\pi} \exp[i(l - m)\Delta\varphi]\, d\Delta\varphi = \begin{cases} 1 & \text{when } l = m \\ 0 & \text{when } l \neq m \end{cases}$$

Hence

$$FT_m(q_i, q_j) = \frac{1}{2\pi} \int_0^{2\pi} \sum_{n_1} \sum_{n_2} I_{n_1}(q_i) I_{n_2}(q_i) I_m(q_j) \exp[i(n_1 + n_2 + m)\varphi] \, d\varphi$$

Also note that

$$\frac{1}{2\pi} \int_0^{2\pi} \exp[i((n_1 + n_2 + m)\varphi] \, d\varphi = \begin{cases} 1 & \text{when } n_1 + n_2 = -m \\ 0 & \text{when } n_1 + n_2 \neq -m \end{cases}$$

So

$$FT_m(q_i, q_j) = \sum_{n_1} I_{n_1}(q_i) I_{-m-n_1}(q_i) I_m(q_j)$$

Let $n_1 = -M$, where $M = \pm 1, \pm 2, \cdots, \pm m_{max}$, except $M = m$

$$FT_m(q_i, q_j) = \sum_M I_{-M}(q_i) I_{M-m}(q_i) I_m(q_j)$$